\documentclass{aa}

\usepackage{txfonts}
\usepackage{graphicx}

\newcommand{\lsi} {LS~I~+61~303}

\newcommand{\prp}    {${\rlap.}^{\prime}$}

\newcommand{\ltsima} {$\; \buildrel < \over \sim \;$}
\newcommand{\simlt}  {\lower.5ex\hbox{\ltsima}}            
\newcommand{\gtsima} {$\; \buildrel > \over \sim \;$}
\newcommand{\simgt}  {\lower.5ex\hbox{\gtsima}}            

\begin{document}

\title{Candidate counterparts to the soft gamma-ray flare in the direction of \lsi}

\author{
A.~J. Mu\~noz-Arjonilla\inst{1,2}
\and J. Mart\'{\i}\inst{1,2}
\and J.~A. Combi\inst{1,5}
\and P. Luque-Escamilla\inst{3,2}
\and J.~R. S\'anchez-Sutil\inst{2}
\and V. Zabalza\inst{4}
\and J.~M. Paredes\inst{4}
}

\offprints{A.~J. Mu\~noz-Arjonilla}

\institute{
Departamento de F\'{\i}sica, EPS,  
Universidad de Ja\'en, Campus Las Lagunillas s/n, Edif. A3, 23071 Ja\'en, Spain \\
\email{ajmunoz@ujaen.es, jmarti@ujaen.es, jcombi@ujaen.es}
\and
Grupo de Investigaci\'on FQM-322, 
Universidad de Ja\'en, Campus Las Lagunillas s/n, Edif. A3, 23071 Ja\'en, Spain \\
\email{jrssutil@ujaen.es}
\and
Dpto. de Ing. Mec\'anica y Minera, EPS,  
Universidad de Ja\'en, Campus Las Lagunillas s/n, Edif. A3, 23071 Ja\'en, Spain \\
\email{peter@ujaen.es}
\and
Departament d'Astronomia i Meteorologia and Institut de Ci\`encies del Cosmos (ICC),
Universitat de Barcelona (UB/IEEC), Mart\'{\i} i Franqu\`es 1, 08028 Barcelona, Spain \\
\email{vzabalza@am.ub.es, jmparedes@ub.edu}
\and
Fac. de Ciencias Astron\'omicas y Geof\'{\i}sicas, 
Universidad Nacional de la Plata, Paseo del Bosque, B1900FWA La Plata, Argentina \\
}

\date{Received / Accepted}

\titlerunning{Counterparts to the SGR flare towards \lsi}

\abstract
{A short duration burst reminiscent of a soft gamma-ray repeater/anomalous X-ray pulsar behaviour was 
detected in the direction of \lsi\ by the \textit {Swift} satellite. While the association
with this well known gamma-ray binary is likely, a different origin cannot be excluded.}
{We explore the error box of this unexpected flaring event and establish the radio, near-infrared and 
X-ray sources in our search for any peculiar alternative counterpart.}
{We carried out a combined analysis of archive Very Large Array radio data of \lsi\ sensitive to
both compact and extended emission. We also reanalysed previous near infrared observations with
the 3.5~m telescope of the Centro Astron\'omico Hispano Alem\'an and X-ray observations with the
\textit {Chandra} satellite.}
{Our deep radio maps of the \lsi\ environment represent a significant advancement on previous work 
and 16 compact radio sources in the \lsi\ vicinity are detected. For some detections, we also identify near infrared 
and X-ray counterparts. Extended emission features in the field are also detected and confirmed. The possible
connection of some of these sources with the observed flaring event is considered. Based on these data, 
we are unable to claim a clear association between the {\it Swift}--BAT flare and any of the sources reported here.
However, this study represents the most sophisticated attempt to determine possible alternative counterparts 
other than \lsi.}


\maketitle

\section {Introduction}

\lsi\ (V615 Cas) is a gamma-ray binary originally discovered in 1977 in the radio
during a survey for variable sources in the Galactic plane (\cite{gregory-78}; \cite{gregory-79}).
The orbital period is about 26.5~d (\cite{hutchings-81}) and it has been detected in data of frequencies 
between radio (\cite{taylor-82}, 1984) and high energy gamma-rays (\cite{albert-08}). 
The physical interpretation of the \lsi\ emission across the complete electromagnetic spectrum
still remains a matter of \mbox{debate} (\cite{romero-07}). Two different models have been proposed 
to explain the full spectral energy distribution: 
(i) a microquasar \mbox{X-ray} binary (\cite{bosch-ramon-06}), and
(ii) a non-accreting pulsar interacting with the envelope of the rapidly rotating Be star (\cite{dubus-06}).

On 2008 September 10, the {\it Swift} Burst Alert Telescope (BAT) detected a burst in the
direction of \lsi\ within its 15--150 keV energy range (\cite{depas-08}). 
The calculated location of this soft gamma-ray short-duration burst-event was found to have 
an error of 2\prp 2 and the position of \lsi\ was found to be clearly consistent with this event (\cite{bart-08}).
Given this coincidence, this burst was associated with magnetar-like activity linked to a young highly magnetized 
pulsar in the binary system (\cite{dubus-atel-08}). Unusual X-ray activity (hard high flux and QPOs) 
had also been reported by \cite{ray-08} just a few weeks before based on PCA data from the {\it RXTE} satellite. 
If the QPOs originate in \lsi, an accretion disk would be necessary to explain the nature of this source.
However, the PCA field of view of $\sim 1$ degree includes many additional sources that could be responsible for 
the QPO behaviour.

Despite these facts, one cannot exclude the possibility of another unrelated source being responsible for 
the observed gamma-ray flare. A population of faint X-ray sources in the vicinity of \lsi\
were reported by \cite{rea-torres-08}, who suggested that one of these sources might be the quiescent counterpart 
of a new transient magnetar. \cite{ma-08} also reported additional X-ray and radio sources coincident with the
{\it Swift}--BAT error circle and some of them with stellar-like counterparts.

We present extensive radio, X-ray, and near infrared observations of the {\it Swift}--BAT
error circle at the location of this 2008 September 10 event. Both archival data and observations
conducted by the authors were used in this paper as described in the log in Table \ref{table-obs}. 
Different populations of sources were detected and their main observational properties are reported. A few 
interesting objects in the direction of the magnetar-like flare are highlighted and 
the possibility of them being alternative candidate counterparts is assessed. The census of radio/X-ray 
sources reported here represents the most complete study to date of alternative counterpart candidates 
to the {\it Swift}--BAT event.

\begin{table}
\caption{\label{table-obs} Radio, near infrared, and X-ray observations of \lsi\ field used in this paper.}

\begin{tabular}{lclcr}
\hline
\hline
         & Instrument    & \multicolumn{1}{c}{Date} & Band & Integ. \\
         &               &             &                        & \mbox{time\ \ } \\
\hline
         &               & 08-Jun-1992  &  C  (6 cm)             & 9.8 h           \\
Radio    &    VLA CnD    & 27-Jun-1992  &  L (20 cm)             & 3.0 h           \\
         &               & 09-Sep-1993* &  C  (6 cm)             & 7.8 h           \\
         &               & 13-Sep-1993* &  C  (6 cm)             & 7.9 h           \\
\hline
         &               & 25-Sep-2007 & $J$\ \ \ (1.25 $\mu$m) & 905 s           \\  
Infrared & CAHA 3.5m     & 25-Sep-2007 & $H$\ \ (1.65 $\mu$m)   & 905 s           \\  
         &               & 25-Sep-2007 & $K_s$ (2.18 $\mu$m)    & 1811 s          \\  
\hline 
X-rays   & {\it Chandra} & 08-Apr-2006 & 0.5--7.0 keV           & 49.9 ks         \\
\hline
\hline 
\end{tabular}
\newline
\newline
(*) Already used by M98.\\
\end{table}

\section {Radio sources within the {\it Swift}--BAT error circle}

The magnetar-like event that renewed interest in studying \lsi\ is probably related to a compact
object formed as a result of a supernova event. Therefore, both compact and extended radio features
could play a role in our understanding of this phenomenon. The environment of \lsi\ in the radio
was studied by \cite{marti-98} (hereafter M98) at 6 cm wavelength using the Very Large Array (VLA) 
of the National Radio Astronomy Observatory (NRAO) with the array in its CnD configuration providing 
appropriate sensitivity to both compact and extended sources. We developed the M98 approach 
improving their sensitivity by combining additional 6 cm VLA archive data acquired by 
the same compact array configuration (see Table \ref{table-obs}) to a total of 25.5~h of on-source time.

\begin{figure}[htpb]
\begin{center}
\vspace{9.5cm}
\includegraphics{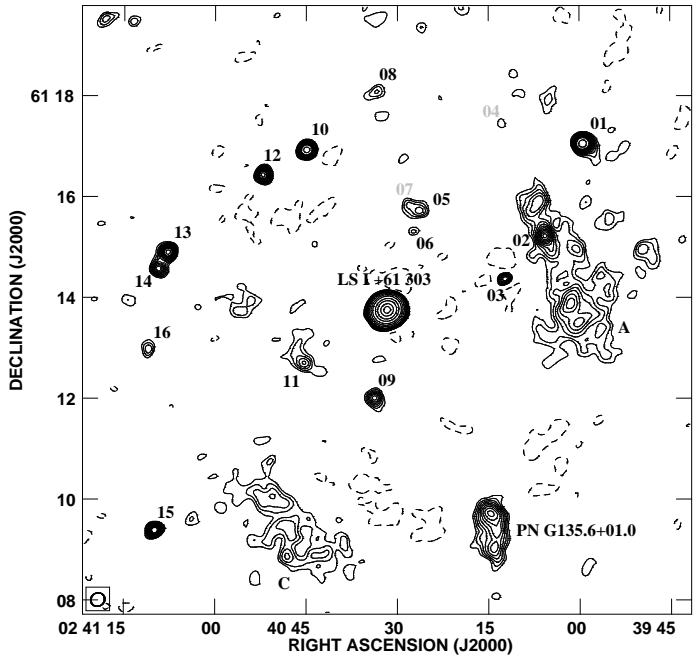}
\caption{Radio map of the \lsi\ environment at the 6 cm wavelength obtained after combining all
VLA observations discussed in the text. The visibility data has been slightly tapered with a 15 k$\lambda$
Gaussian in order to enhance the extended emission in the field. The X-ray binary \lsi\ and the nearby planetary
nebula G135.6+01.0 \cite{bond2003} are labelled and the other relevant sources numbered in right ascension order. Radio sources marked
with grey numbers are better seen in the non-tapered map of Fig. \ref{circle_maps}.
The rms noise is at the level of 14 $\mu$Jy beam$^{-1}$. The beam size is shown at the
bottom left corner and corresponds to $17^{\prime \prime}.0 \times 15^{\prime \prime}$.7,
with position angle of $-53^{\circ}.3$.
Contours shown are $-3$, 3, 4, 5, 6, 7, 8, 9, 10, 11, 12, 13, 15, 20, 25, 50, 100, 200, 500, 1000, 2000 and 3000 
times the rms noise.}
\label{vla6cm_taper}
\end{center} 
\end{figure}

\begin{figure*}[htpb]
\begin{center}
\vspace{9.5cm}
\includegraphics{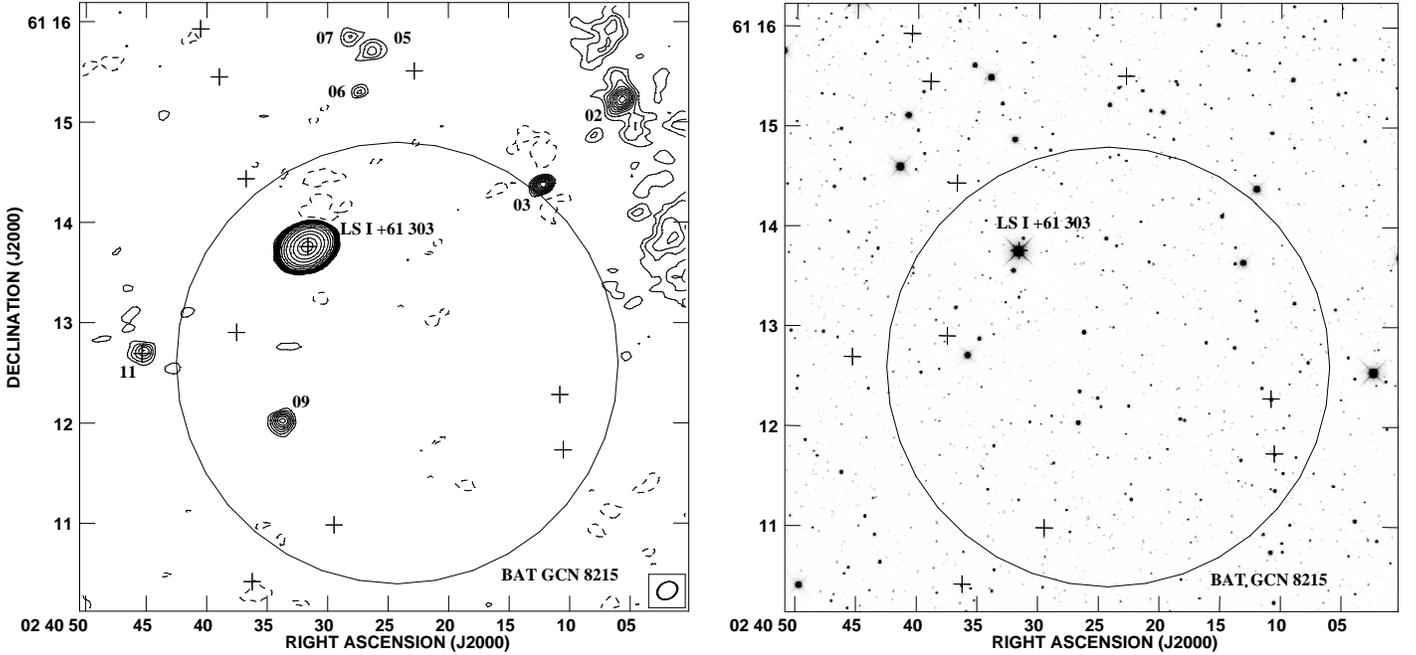}
\caption{{\bf Left.} Zoomed contour radio map of the \lsi\ environment at 6 cm after combining all the 
VLA data used in this paper. Natural weight with no taper has been used. The big circle corresponds to 
the refined 90\% confidence BAT location for the magnetar-like flare in the direction of \lsi. 
The rms noise is at the level of 13 $\mu$Jy beam$^{-1}$. The beam size is shown at the bottom right 
corner and corresponds to $12^{\prime \prime}.7 \times 9^{\prime \prime}$.9, with position angle of 
$-58^{\circ}.8$. Contours shown are $-3$, 3, 4, 5, 6, 7, 8, 9, 10, 11, 12, 13, 15, 20, 25, 50, 100, 200, 
500, 1000, 2000 and 3000  times the rms noise. Crosses show the X-ray {\it Chandra} ACIS-I sources with 
a signal-to-noise ratio above 3. 
{\bf Right.} The same field as observed in the near infrared $K_s$-band with the CAHA 3.5~m telescope 
and the OMEGA2000 camera.}
\label{circle_maps}
\end{center}
\end{figure*}

The AIPS package of NRAO was used for the standard interferometer data processing and self-calibration.
We removed the variable \lsi\ core to avoid artifacts and replaced it for cosmetic reasons with a constant 
point source component with the observed average flux density. We also removed a nearby bright radio source, 
whose presence affected significantly the dynamic range of the maps.
The final result is presented in  Fig. \ref{vla6cm_taper}, where extended emission is enhanced with a
slight taper. Compact sources are more clearly evident in the non-tapered radio map of Fig. \ref{circle_maps}
with a rms noise of 13 $\mu$Jy beam$^{-1}$, significantly lower than that of M98. The observational
properties of these compact radio sources are listed in Table \ref{table-sources}, where J2000.0 positions
are given.

\section {Near infrared observations of the {\it Swift}--BAT error circle}

The field of \lsi\ was observed at near infrared wavelengths with the 3.5 m telescope
at the Centro Astron\'omico Hispano Alem\'an (CAHA) in Almer\'{\i}a (Spain), one year before the {\it Swift} event.
The OMEGA2000 camera was used and the images were taken through the $J$, $H$ and $K_s$ filters. 
CAHA observations were processed following the standard procedures
for sky background subtraction, flat-fielding and median-combining of individual frames using
the IRAF\footnote{$<$http://iraf.noao.edu/$>$} software package. Astrometry in the final frames was determined by identifying
about twenty stars in the field for which positions were retrieved from the 2MASS catalog.
The relevant part of the resulting image is shown in the right panel of Fig. \ref{circle_maps}.
Photometric and astrometric data derived from this image are included in Table \ref{table-sources}
for VLA radio sources with a near infrared counterpart. The errors in the photometric measurements 
are dominated by the uncertainty in the zero point, which is estimated to be about 0.1 mag in all filters.
The existence of a candidate counterpart was assessed based on the statistical parameter $r$ 
(\cite{allington-smith-82}) defined to be:
\begin{equation} 
r = \sqrt{\frac{(\Delta \alpha \cos \delta)^2}{\sigma_{\alpha \rm{,IR}}^2 + \sigma_{\alpha \rm{,rad}}^2} +
\frac{(\Delta \delta)^2}{\sigma_{\delta \rm{,IR}}^2 + \sigma_{\delta \rm{,rad}}^2}}  
\end{equation}
\noindent where $\Delta \alpha$ and $\Delta \delta$ are the differences between the measured
near infrared and radio positions, and $\sigma_{\alpha \rm{,IR}}$, $\sigma_{\alpha \rm{,rad}}$,
$\sigma_{\delta \rm{,IR}}$ and $\sigma_{\delta \rm{,rad}}$ are the infrared and radio uncertainties
in right ascension and declination.

The offsets between radio and near infrared positions are also listed in Table \ref{table-sources}.
A value of $r\leq 3$ is taken to be indicative of astrometric coincidence
within the combined radio and near infrared position errors.

\begin{table*}
\caption{Compact radio sources in the field around \lsi\ and their candidate near infrared counterparts}
\label{table-sources}
\begin{center}
\begin{tabular}{lccccrrrrrr}
\hline
\hline
VLA  & Right Ascension & Declination & Peak flux dens. & Total flux dens. & \multicolumn{1}{c}{$\Delta\alpha \cos\delta$} & \multicolumn{1}{c}{$\Delta\delta$} &
\multicolumn{1}{c}{$r$} & \multicolumn{1}{c}{$J$} & \multicolumn{1}{c}{$H$} & \multicolumn{1}{c}{$K_s$} \\
Id. \#    &  (hms) & ($^{\circ}$\ $^{\prime}$\ $^{\prime\prime}$)  & (mJy beam$^{-1})$ & (mJy) & \multicolumn{1}{c}{($^{\prime\prime}$)} & \multicolumn{1}{c}{($^{\prime\prime}$)} &  & \multicolumn{1}{c}{(mag)} & \multicolumn{1}{c}{(mag)} & \multicolumn{1}{c}{(mag)} \\
\hline
01*$^{\rm (a)}$  & 02 39 59.44(0.04) & +61 17 03.3(0.2) & 0.87(0.04) & 1.00(0.07) & +0.43 & +0.00 &  0.71 & 20.1 & 18.8 & 18.1 \\
02   & 02 40 05.63(0.13) & +61 15 13.5(1.0) & 0.23(0.04) & 0.41(0.09) &  $-$  &  $-$  &  $-$  & $-$  & $-$  & $-$  \\
03*  & 02 40 12.22(0.11) & +61 14 22.4(0.7) & 0.20(0.04) & 0.15(0.05) & -1.01 & +0.70 &  1.17 & 11.4 & 10.3 & 10.0 \\
04   & 02 40 12.77(0.24) & +61 17 25.9(1.4) & 0.11(0.04) & 0.10(0.06) & -0.94 & -0.10 &  0.27 & $-$  & 18.8 & 17.7 \\
05   & 02 40 26.32(0.47) & +61 15 43.2(2.4) & 0.08(0.04) & 0.17(0.10) & +0.50 & -0.90 &  0.38 & 18.0 & 17.1 & 16.0 \\
06   & 02 40 27.34(0.34) & +61 15 18.6(1.9) & 0.07(0.04) & 0.06(0.06) & +0.43 & -0.10 &  0.10 & $-$  & $-$  & 18.2 \\
07   & 02 40 28.13(0.43) & +61 15 50.8(2.6) & 0.07(0.04) & 0.12(0.09) & +4.47 & +3.80 &  1.62 & 15.0 & 14.5 & 14.3 \\
08   & 02 40 33.33(0.26) & +61 18 05.8(1.9) & 0.12(0.04) & 0.24(0.10) &  $-$  &  $-$  &  $-$  & $-$  & $-$  & $-$  \\ 
09*  & 02 40 33.74(0.20) & +61 12 00.8(1.4) & 0.14(0.04) & 0.17(0.07) & -0.36 & -0.70 &  0.51 & 19.2 & 17.8 & 16.7 \\
10*$^{\rm (a, b)}$  & 02 40 44.94(0.05) & +61 16 55.8(0.3) & 0.54(0.04) & 0.49(0.06) & +0.07 & +0.20 &  0.65 &  6.4 &  6.2 &  6.2 \\
11$^{\rm (a)}$   & 02 40 45.32(0.33) & +61 12 42.3(2.0) & 0.10(0.04) & 0.17(0.09) &  $-$  &  $-$  &  $-$  & $-$  & $-$  & $-$  \\
12*  & 02 40 52.15(0.06) & +61 16 26.1(0.4) & 0.39(0.04) & 0.35(0.06) & +0.43 & +0.10 &  0.54 & $-$  & $-$  & 18.1 \\
13*$^{\rm (a)}$  & 02 41 07.76(0.06) & +61 14 54.2(0.4) & 0.46(0.04) & 0.49(0.07) & +0.79 & +0.30 &  1.14 & 20.5 & 19.6 & 18.2 \\
14*  & 02 41 09.36(0.07) & +61 14 34.5(0.5) & 0.34(0.04) & 0.34(0.07) &  $-$  &  $-$  &  $-$  & $-$  & $-$  & $-$  \\
15*  & 02 41 09.94(0.03) & +61 09 23.0(0.2) & 1.16(0.04) & 1.00(0.06) &  $-$  &  $-$  &  $-$  & $-$  & $-$  & $-$  \\
16*  & 02 41 10.97(0.14) & +61 12 58.1(1.2) & 0.18(0.04) & 0.23(0.08) &  $-$  &  $-$  &  $-$  & $-$  & $-$  & $-$  \\
\hline
\hline
\end{tabular}
\end{center}
(*) Sources already detected by M98.\\ 
$^{\rm (a)}$ Candidate X-ray counterpart found for this source.\\
$^{\rm (b)}$ Near infrared magnitudes taken from the 2MASS All-Sky Catalog of point sources.
\end{table*}

\section {X-ray sources within the {\it Swift}--BAT error circle}

\begin{table*}
\centering
\caption{X-ray sources with a signal-to-noise ratio above $3\sigma$ in the \lsi\ vicinity.}
\label{table-x-rays}
\begin{tabular}{ccccc}
\hline
\hline
{\it Chandra} & Right Ascension & Declination     & Flux        & Hardness \\
Id. \# & (hms) & ($^{\circ}$\ $^{\prime}$\ $^{\prime\prime}$) & ($10^{-6}$ ph cm$^{-2}$ s$^{-1}$) & Ratio \\
\hline
01 & 02 39 54.899(0.024) & +61 12 39.80(0.24) &  0.66(0.27) &  0.82(0.08) \\
02 & 02 39 58.962(0.019) & +61 15 19.46(0.13) &  2.44(0.43) & -0.93(0.35) \\
03 & 02 39 59.445(0.016) & +61 17 03.53(0.10) &  3.98(0.53) &  0.33(0.09) \\
04 & 02 40 01.019(0.024) & +61 16 45.73(0.14) &  2.43(0.43) & -0.49(0.28) \\
05 & 02 40 02.592(0.023) & +61 17 28.46(0.16) &  2.69(0.48) &  0.21(0.15) \\
06 & 02 40 10.328(0.036) & +61 17 11.64(0.10) &  1.09(0.31) & -0.48(0.46) \\
07 & 02 40 10.521(0.017) & +61 11 44.04(0.13) &  0.77(0.27) & -0.86(0.70) \\
08 & 02 40 10.804(0.014) & +61 12 16.98(0.13) &  1.01(0.29) & -0.87(0.57) \\
09 & 02 40 16.604(0.018) & +61 17 37.79(0.18) &  1.22(0.34) & -0.44(0.43) \\
10 & 02 40 22.166(0.030) & +61 17 57.30(0.16) &  1.09(0.33) & -0.27(0.42) \\
11 & 02 40 22.802(0.018) & +61 08 47.61(0.26) &  0.83(0.29) & -1.13(0.80) \\
12 & 02 40 22.821(0.007) & +61 15 30.59(0.05) &  2.42(0.40) &  0.02(0.17) \\
13 & 02 40 24.507(0.015) & +61 17 20.90(0.13) &  1.75(0.36) &  0.59(0.09) \\
14 & 02 40 26.729(0.011) & +61 16 20.20(0.13) &  1.37(0.33) & -0.54(0.39) \\
15 & 02 40 26.859(0.016) & +61 16 29.57(0.15) &  1.36(0.33) &  0.28(0.19) \\
16 & 02 40 28.861(0.011) & +61 16 43.71(0.09) &  3.06(0.48) & -0.44(0.24) \\
17 & 02 40 29.471(0.011) & +61 10 59.21(0.11) &  0.66(0.24) &  0.91(0.04) \\
18 & 02 40 36.234(0.017) & +61 10 25.18(0.14) &  0.70(0.25) & -0.66(0.65) \\
19 & 02 40 36.754(0.015) & +61 14 26.11(0.14) &  0.77(0.30) &  0.44(0.24) \\
20 & 02 40 37.526(0.009) & +61 12 54.33(0.19) &  0.53(0.22) & -0.25(0.58) \\
21 & 02 40 38.956(0.006) & +61 15 27.01(0.04) &  6.22(0.62) &  0.24(0.08) \\
22 & 02 40 40.505(0.028) & +61 15 55.76(0.14) &  0.48(0.25) & -0.06(0.61) \\
23 & 02 40 44.029(0.019) & +61 16 54.28(0.13) &  1.82(0.41) & -0.59(0.38) \\
24 & 02 40 44.233(0.040) & +61 18 16.45(0.19) &  0.97(0.33) &  0.02(0.37) \\
25 & 02 40 44.384(0.022) & +61 17 28.50(0.09) &  4.70(0.58) & -0.45(0.18) \\
26 & 02 40 44.944(0.004) & +61 16 56.10(0.02) & 36.61(1.51) & -0.81(0.08) \\
27 & 02 40 45.346(0.019) & +61 12 41.61(0.12) &  0.51(0.22) &  0.45(0.26) \\
28 & 02 40 47.666(0.013) & +61 16 17.56(0.06) &  3.05(0.76) & -0.24(0.33) \\
29 & 02 40 51.254(0.007) & +61 14 28.08(0.05) &  4.21(0.52) & -0.61(0.21) \\
30 & 02 40 53.383(0.024) & +61 15 55.31(0.09) &  0.88(0.28) &  0.43(0.20) \\
31 & 02 40 53.754(0.014) & +61 14 29.60(0.11) &  0.61(0.24) & -0.11(0.48) \\
32 & 02 40 59.824(0.032) & +61 09 31.73(0.25) &  0.82(0.30) & -0.63(0.63) \\
33 & 02 41 00.953(0.018) & +61 11 14.55(0.16) &  2.56(0.51) & -0.15(0.24) \\
34 & 02 41 02.787(0.010) & +61 14 00.39(0.07) &  4.14(0.52) &  0.58(0.06) \\
35 & 02 41 04.715(0.036) & +61 15 32.62(0.23) &  1.85(0.43) &  0.08(0.23) \\
36 & 02 41 07.025(0.033) & +61 16 28.63(0.20) &  1.28(0.36) & -0.11(0.33) \\
37 & 02 41 07.835(0.021) & +61 14 54.69(0.14) &  3.98(0.55) &  0.24(0.11) \\
38 & 02 41 08.254(0.027) & +61 08 06.78(0.23) &  4.82(0.65) & -0.86(0.26) \\
39 & 02 41 10.223(0.019) & +61 16 08.34(0.13) &  5.51(0.63) &  0.55(0.05) \\
40 & 02 41 14.310(0.043) & +61 16 30.62(0.21) &  1.29(0.36) &  0.58(0.12) \\
\hline
\hline
\end{tabular}
\end{table*}

We also used X-ray data from an observation of the \lsi\ field obtained with {\it Chandra} about two years before 
the {\it Swift} event using the standard ACIS-I setup in VF mode during a total of 49.9~ks. 
The data were reduced using the {\it Chandra} Interactive Analysis of Observations software package (CIAO v4.0) 
and CALDB v3.4.5. The analysis of \lsi\ data itself was discussed and its results reported in detail by
\cite{paredes-07}. We used the CIAO tool \verb+wavdetect+ to obtain a list of candidate sources with data of a 
signal-to-noise ratio above $3\sigma$. Most of the sources were very faint and the counts insufficient
for a robust spectral analysis, so we obtained an exposure map of the
observation and derived the photon fluxes and hardness ratios of the detected sources.
In Table \ref{table-x-rays}, we present the fluxes in the \mbox{0.5--7.0 keV} band and the hardness ratio,
defined as $(H-S)/(H+S)$, where $H$ is the flux in the 2.0--7.0 keV band and $S$ is the flux in the
0.5--2.0 keV band.

The X-ray observation completely covers the area observed in radio with the VLA.
In this region, we found a total of 40 sources, excluding \lsi\ itself, 39 of which were previously
unknown.

\section {Discussion and conclusions}

Our deep radio maps in Figs. \ref{vla6cm_taper} and \ref{circle_maps} exhibit 16 compact
radio sources in addition to \lsi\ with a peak flux density of a factor of four above the rms noise.
When we restricted ourselves to the {\it Swift}--BAT error circle, three compact radio sources
(labelled as \# 03, 09 and 11 in Table \ref{table-sources}) appear to be located within or close to the 
refined BAT location. We measured their flux densities at 6 cm at the three epochs quoted in 
Table \ref{table-obs} separately, and no evidence of variability was found in any source.
The VLA sources \# 03 and 09 are consistent with point-like near infrared counterparts
(see Table \ref{table-sources}). Source \# 03 coincides with a particularly bright object ($K_s=10.0$) 
and its observed colours (e.~g. $J-K_s \simeq 1.4$) are reminiscent of a giant star with a late spectral 
type, provided that interstellar extinction is not too high. Source \# 09 appears to be a highly reddened object 
with $J-K_s \simeq 2.5$ mag and its stellar or extragalactic nature cannot be established based on our photometry alone.
Neither of them is detected in our {\it Chandra} observations. On the other hand and using the same Eq. \# (1) criterion, 
VLA source \# 11 is coincident with a {\it Chandra} X-ray source (labelled as \# 27 in Table 
\ref{table-x-rays}) but no infrared counterpart is detected.
If either the magnetar-like burst behaviour or the X-ray activity presenting QPOs are unrelated to \lsi, then 
these radio sources appear to be potential alternative counterparts for these detections.

We note that VLA source \# 11 is located at the centre of the extended radio source D
($\alpha = 02^h 40^m 45^s$, $\delta = +61^{\circ} 12{\rlap.}^{\prime}7$)
reported by M98 which was proposed to be a possible large-scale lobe powered by \lsi.
In our new map (Fig. \ref{vla6cm_taper}), this object resembles more closely a background radio galaxy
with bent bipolar jets. The X-ray detection might then originate in the central core of the radio galaxy.
In this interpretation, the connection of this source to a magnetar-like flare/QPOs appears unlikely
since this kind of behaviour is not typical of an extragalactic object.

The new VLA maps confirm all the extended radio emission features reported by M98.
A possible supernova remnant origin was tentatively considered by these authors but the
low surface brightness and absence of extended X-rays rendered it unlikely. 
Revisiting this scenario seems timely because the magnetar-like emitter, even if it is not \lsi, 
is naturally expected to be a galactic compact object created during a past supernova event. 
Thus, we attempted to estimate the spectral index of the two brightest extended radio features
located west ($\alpha = 02^h 40^m 01^s$, $\delta = +61^{\circ} 13{\rlap.}^{\prime}9$)
and south ($\alpha = 02^h 40^m 47^s$, $\delta = +61^{\circ} 09{\rlap.}^{\prime}5$)
of \lsi\ in Fig. \ref{vla6cm_taper} (sources A and C in M98) by combining our 6 cm map with the 20 cm
archive data also quoted in Table \ref{table-obs}.
In this process, the visibility data was appropriately constrained to match angular resolutions in both data sets.
The resulting spectral index map is not conclusive enough due to the poor quality of the 20 cm data.
Only the brightest parts of the western radio feature have a spectral index error below 0.1
suggesting a possible non-thermal origin ($S_\nu \propto \nu^{-0.7}$). However, {\it Chandra} observations
reveal no X-ray counterpart, even for this western extended radio feature. Improved long wavelength radio and X-ray
observations are required to constrain more accurately the possibility of a faint supernova remnant associated with \lsi\ 
or any other source in the field. Concerning compact sources, only one source labelled as \# 15 was detected at 20 cm.
Its spectral index ($-0.6 \pm 0.2$) may suggest a non-thermal emission mechanism for this source.

In conclusion, we have reported a handful of X-ray and radio sources within or close to the improved
{\it Swift}--BAT location of the magnetar-like and QPO events towards \lsi . No object displays
any peculiar signature that could reveal a possible connection with these unusual phenomena. Although we cannot exclude them 
being alternative counterpart candidates, the most likely possibility is that
\lsi\ is actually behind the observed flaring event.

\begin{acknowledgements}
{\small The authors acknowledge support by
grants AYA2007-68034-C03-02 and AYA2007-68034-C03-01
from the Spanish government, and FEDER funds.
This has been also supported by Plan Andaluz de Investigaci\'on
of Junta de Andaluc\'{\i}a as research group FQM322.
The NRAO is a facility of the NSF
operated under cooperative agreement by Associated Universities, Inc.
This paper is also based on observations collected at the Centro Astron\'omico Hispano Alem\'an
(CAHA) at Calar Alto, operated jointly by the Max-Planck Institut
f\"ur Astronomie and the Instituto de Astrof\'{\i}sica de Andaluc\'{\i}a (CSIC).
This research made use of the SIMBAD
database, operated at the CDS, Strasbourg, France.
}
\end{acknowledgements}

\end{document}